\documentclass[aps,prl,twocolumn,superscriptaddress,showpacs,longbibliography]{revtex4-1}
\usepackage{amsfonts,amsmath,bm}
\usepackage{mathtools}
\usepackage{graphicx}
\usepackage{epstopdf}

\DeclarePairedDelimiter\norm{\lVert}{\rVert}%
\DeclarePairedDelimiter\abs{\lvert}{\rvert}%
\makeatletter
\let\oldabs\abs
\def\abs{\@ifstar{\oldabs}{\oldabs*}}
\let\oldnorm\norm
\def\norm{\@ifstar{\oldnorm}{\oldnorm*}}
\makeatother

\begin{document}
\title{Coherent phonon lasing in a thermal quantum nanomachine}  

\author{P. Karwat}
\affiliation{Department of Theoretical Physics, Wroc\l{}aw University of Science and Technology, Wybrze\.ze Wyspia\'nskiego 27, 50-370 Wroc\l{}aw, Poland\\}
\author{D. E. Reiter}
\affiliation{Institut f\"ur Festk\"orpertheorie, Westf\"alische Wilhelms-Universit\"at M\"unster, Wilhelm-Klemm-Strasse 10, 48149 M\"unster, Germany \\}
\affiliation{The Blackett Laboratory, Department of Physics, Imperial College London, South Kensington Campus, SW7 2AZ, London, United Kingdom \\} 
\author{T. Kuhn}
\affiliation{Institut f\"ur Festk\"orpertheorie, Westf\"alische Wilhelms-Universit\"at M\"unster, Wilhelm-Klemm-Strasse 10, 48149 M\"unster, Germany \\}
\author{O. Hess}
\affiliation{The Blackett Laboratory, Department of Physics, Imperial College London, South Kensington Campus, SW7 2AZ, London, United Kingdom \\} 

\begin{abstract}
The notion of nanomachines has recently emerged to engage and use collective action of ensembles of nanoscale components or systems. Here we present a heat-gradient driven nanomachine concept which through appropriate coupling between quantum nanosystems is capable of realising and maintaining an inversion. Based on a Lindblad form of the Quantum Master Equation with a semiclassical coupling to the lattice displacement phonon field we show that this positive inversion can be harnessed to generate coherent optomechanical oscillations and phonon lasing. 
\end{abstract}


\maketitle

% \% Put \label in argument of \section for cross-referencing

%TEXT FOR INTRODUCTION
In nature, an inversion of the occupation probabilities of quantum states is clearly quite artificial since the standard thermal situation leads to the opposite - to a Boltzmann distribution over the energy levels where lower levels are populated exponentially larger than higher ones. Typically such an inversion is achieved by pumping the system, e.g. electrically or optically. Once achieved, an inversion will allow that an initially very small field is amplified by several orders of magnitude resulting in lasing. This leads to the question: Could it be possible to use a temperature difference not only to drive a steam engine, but as source for lasing? 

In this letter, we will tackle this question by considering a nanomachine that emits phonons and study, if applying a heat gradient results in an amplification of the emission, i.e., to phonon lasing. The concept of a nanomachine, that emits phonons rather than photons, has been on the minds of many researchers \cite{grudinin10, beardsley10, carmele17, vahala09, mendona10, kepesidis13, mahboob13}. In such a system it is interesting to exploit the possibility of amplification of the phonon wave, which in analogue to lasing could lead to the construction of a phonon-laser or saser \cite{zavtrak1997saser,tilstra2007optically}. Such a phonon-laser could be used for new type highly precise nondestructive measurements \cite{khurgin10}. Several proposal for phonon lasing haven been put forward \cite{mahboob13,vahala09,sander12,kent10,vahala10} and also first implementation of phonon lasing using quantum-well structures have been reported \cite{maryam2013dynamics}. In contrast to these studies, in our proposal we want to achieve phonon lasing in a nanomachine using a heat gradient.

\begin{figure}[t]
	\centering
	\includegraphics[width=\columnwidth]{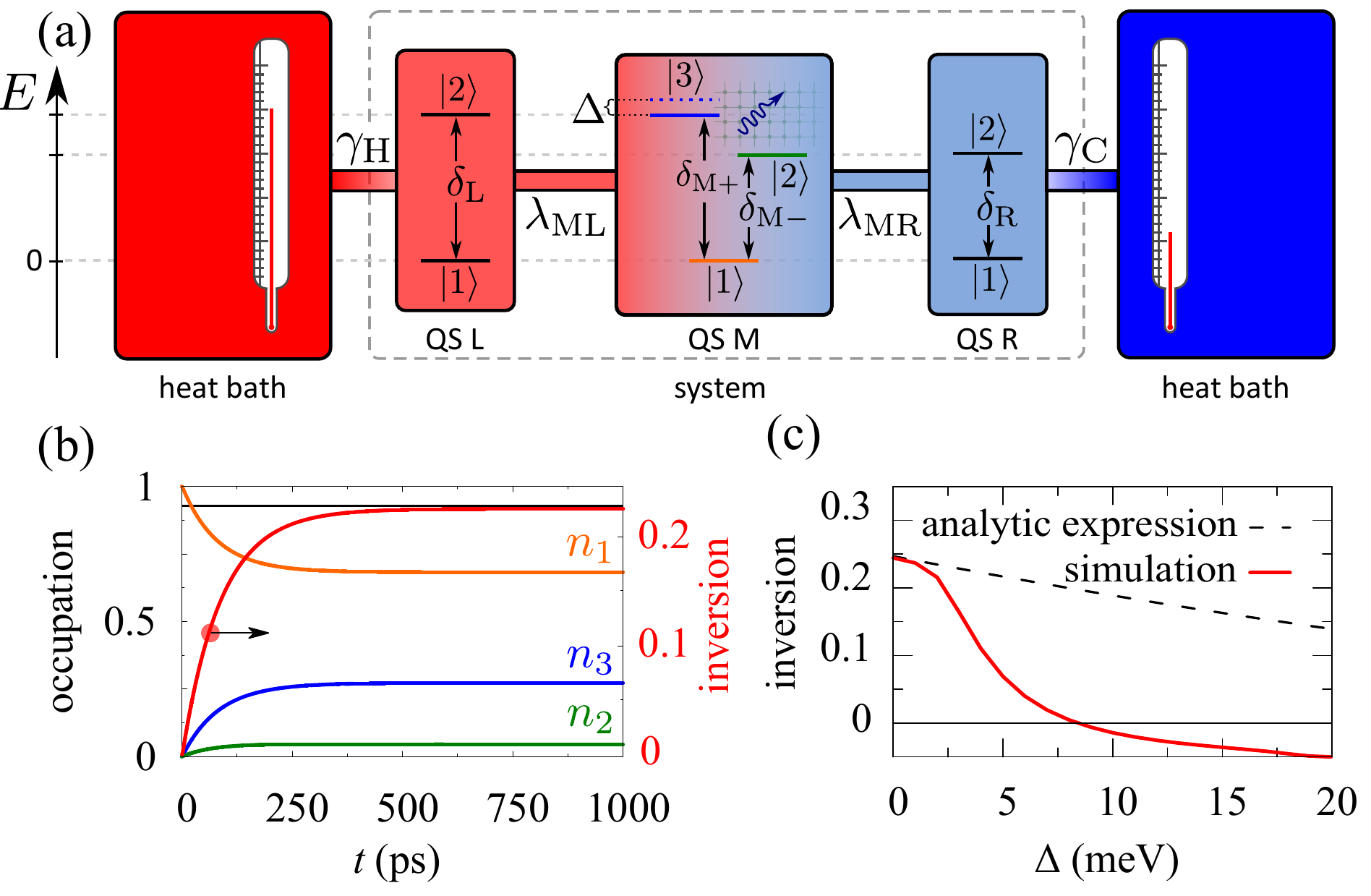}
	\caption{(a) Sketch of the system. (b) Occupations $n_i$ of the middle QS as function of time for temperatures $T_{\mathrm{H}}=400$~K and $T_{\mathrm{C}}=100~K$ as well as the inversion $I=n_3-n_2$. (c) Inversion as function of energy mismatch $\Delta$. The dashed lines marks the ideal inversion  from Eq.~\eqref{eq:idealI}.
	\label{fig:model_system}}
\end{figure}

%\textbf{System set-up:}\\
Our nanomachine is composed of three coupled quantum systems (QSs), which are subject to a heat gradient as displayed in Fig.~\ref{fig:model_system}(a). The active medium is given by the middle three-level system (QS~M). The central quantum system interacts with two-level subsystems (QS~L/R) at each side, which act as energy filters. Such filtering is necessary, because a direct coupling to the two heat baths with different temperatures would lead to a thermal occupation of the system and not to inversion. Each filter is coupled to a heat bath, where the left bath has a significantly higher temperature ($T_{\mathrm{H}}$) than the right one ($T_{\mathrm{C}}$). In consequence of the temperature difference, a flow of excitation takes place. We will show that the exclusive thermalization of the resonant transitions may lead to the crucial inversion in the upper two levels of the central system for certain parameters. The inversion can then result in the emission of coherent phonons at the central quantum system. 

The Hilbert space of the nanomachine is spanned by the product of the basis states in the QSs as depicted in Fig.~\ref{fig:model_system}(a). The relative energies of the states in the two-level systems are parametrized by $\delta_{\mathrm{L/R}}$ and in the central three-level system by $\delta_{\mathrm{M_{+}/M_{-}}}$. The system Hamiltonian reads 
\begin{equation}
	\label{ham_loc}
	\hat{H}_{\mathrm{sys}} = \hat{h}^{(\mathrm{L})} + \hat{h}^{(\mathrm{M})} + \hat{h}^{(\mathrm{R})},
\end{equation}
where $\hat{h}^{(\mathrm{L})/(\mathrm{M})/(\mathrm{R})}$ are the Hamiltonians of the QSs with
\begin{equation}
	\label{ham_L}
	\hat{h}^{\mathrm{(L)}} = \sum_{n}\epsilon_{n}^{(\mathrm{L})}P_{nn}^{\mathrm{(L)}}\otimes \hat{1}^{\mathrm{(M)}}\otimes\hat{1}^{(\mathrm{R})} \,.
\end{equation}
Here, $\epsilon^{(\mathrm{L})}_{n}$ are the energies, $\hat{P}^{(\mathrm{L})}_{nm} = |n\rangle^{(\mathrm{L})} {}^{(\mathrm{L})} \langle m|$ is the projection operator and $\hat{1}^{(\mathrm{M})/(\mathrm{R})}$ denote unity operators in the space of respective subsystem. The Hamiltonians $\hat{h}^{(\mathrm{M})}$ and $\hat{h}^{(\mathrm{R})}$ are constructed in the same way.

The QSs are coupled in a way that excitations can be exchanged between adjacent sites as denoted in Fig.~\ref{fig:model_system}(a), i.e. the middle system interacts with both sides, while the interaction between left and right system is suppressed. The corresponding Hamiltonian reads
\begin{eqnarray}
	\label{ham_int}
	\hat{H}_{\mathrm{int}}= \lambda_{\mathrm{ML}}\bigg( \hat{h}^{(\mathrm{LM})}\otimes \hat{1}^{\mathrm{(R)}} \bigg) 
		+\lambda_{\mathrm{MR}}\bigg(\hat{1}^{\mathrm{(L)}}\otimes  \hat{h}^{(\mathrm{MR})} \bigg)
\end{eqnarray}
where $\lambda_{\mathrm{ML/MR}}$ is the coupling parameter. The coupling
\begin{equation}
\hat{h}^{(\mathrm{LM})} = \hat{P}_{21}^{\mathrm{(L)}} \otimes \bigg( \hat{P}_{12}^{\mathrm{(M)}} + \hat{P}_{13}^{\mathrm{(M)}} + \hat{P}_{23}^{\mathrm{(M)}} \bigg) + \mathrm{h.c.}
\end{equation}
is given by the respective projection operators and analogous for $\hat{h}^{(\mathrm{MR})}$. The coupling is taken to be weak such that the energy contribution of the interaction is small compared to the energy contained in the system. 

%\textbf{Heat coupling- Theory}\\
Each of the two edge QSs is coupled locally to a heat bath of different temperature. To describe the coupling, we make use of a Quantum Master Equation within a Lindblad form, which accounts for the non-equilibrium situation in our system \cite{breuer02}. For this, we set up the equation of motion for the density matrix $\hat{\rho}$ via
\begin{equation}
	\frac{d\hat{\rho}}{dt} = -\frac{i}{\hbar}[\hat{H}_{\mathrm{sys}}+\hat{H}_{\mathrm{int}},\hat{\rho}] + \hat{D}_{\mathrm{H}}(\hat{\rho}) + \hat{D}_{\mathrm{C}}(\hat{\rho})\,.
\end{equation}
$\hat{D}_{\mathrm{H}}$ and $\hat{D}_{\mathrm{C}}$ are the dissipators to the hot (left) and cold (right) heat bath, respectively,
\begin{equation}
	\hat{D}_{\mathrm{H}}(\hat{\rho}) = \sum_{k = 1}^{2}\Gamma_{k}(T_{\mathrm{H}})\bigg(\hat{L}^{(\mathrm{L})}_{k}\hat{\rho}\hat{L}^{(\mathrm{L})\dagger}_{k} 
	- \frac{1}{2}[\hat{L}^{(\mathrm{L})\dagger}_{k}\hat{L}^{(\mathrm{L})}_{k}, \hat{\rho}]_{+} \bigg)
\end{equation}
with the Lindblad operators 
\begin{equation*}
	\hat{L}^{(\mathrm{L})}_{1} = \hat{P}^{(\mathrm{L})}_{21}\otimes\hat{1}^{(\mathrm{M,R})},\qquad
	\hat{L}^{(\mathrm{L})}_{2} = \hat{P}^{(\mathrm{L})}_{12}\otimes\hat{1}^{(\mathrm{M,R})} \, .
\end{equation*}
The effectiveness of the heat coupling is given by the rates $\Gamma_{k}$ chosen by a phenomenological ansatz for the spectral density of an environment of Ohmic kind \cite{breuer02} with
\begin{equation}
	\Gamma_{k}(T_{\mathrm{H}}) = \frac{\gamma}{1+\exp \big\{(-1)^{k-1}\delta_{\mathrm{L}}/k_{\mathrm{B}}T_{\mathrm{H}}\big\}}.
\end{equation}
containing the distribution function. The dissipator describing the coupling to the cold (right) heat bath  $\hat{D}_{\mathrm{C}}(\hat{\rho})$ is analogue. \\
 Our goal is to achieve an inversion between states $|3\rangle ^{(\mathrm{M})}$ and $|2\rangle ^{(\mathrm{M})}$, which later will be coupled to a phonon mode. The energy of typical acoustic phonons lie in the order of a few meV. Setting the energy of the lower state $|1\rangle^{(\mathrm{M})}$ to zero, we accordingly chose $\delta_{\mathrm{M_{+}}} = 30$~meV and $\delta_{\mathrm{M_{-}}} = 25$~meV, thus $\delta_{\mathrm{M_{+}}} - \delta_{\mathrm{M_{-}}}= 5$~meV. The energies of the edge state are set to $\delta_{\mathrm{L}} = \delta_{\mathrm{M_{+}}}=30$~meV  and $\delta_{\mathrm{R}} = \delta_{\mathrm{M_{-}}}=25$~meV. If not stated otherwise, the parameters are set to $\lambda=\lambda_{\mathrm{ML}} = \lambda_{\mathrm{MR}} = 0.03$~meV, $\gamma=\gamma_{\mathrm{H}} = \gamma_{\mathrm{C}} = 3$~ps$^{-1}$. As initial condition we assume that the whole system is in its ground state $|1\rangle ^{(\mathrm{L})}\otimes |1\rangle ^{(\mathrm{M})} \otimes|1\rangle ^{(\mathrm{R})}$. 

%\textbf{Inversion - Results}\\
By solving the equation of motion we calculate the occupations $n_i = \langle P^{(\mathrm{M})}_{ii}\rangle $ of the three states in QS~M. The time evolution of the occupations for a hot bath with temperature $T_{\mathrm{H}}=400$~K and a cold bath of $T_{\mathrm{C}}=100$~K is shown in Fig.~\ref{fig:model_system}(b). When the heating gradient is switched on at $t=0$, the occupation $n_1$ decreases in favour of $n_2$ and $n_3$. After a few hundreds of picoseconds a stationary state is reached. Due to the heat gradient and the energy filtering an inversion in the central system is achieved (red solid line). It is interesting to compare the achieved inversion to the ideal case in which the occupations $n_{i}$ will follow the Boltzmann distribution
\begin{equation}
\label{P3toP1}
	\frac{n_{3}}{n_{1}} = \exp\bigg\{-\frac{\delta_{\mathrm{M+}}}{k_{\mathrm{B}}T_{\mathrm{H}}}\bigg\}, \qquad
	\frac{n_{2}}{n_{1}} = \exp\bigg\{-\frac{\delta_{\mathrm{M-}}}{k_{\mathrm{B}}T_{\mathrm{C}}}\bigg\}.
\end{equation}
The inversion in the ideal case is then
\begin{equation}
\label{eq:idealI}
	I=n_{3} - n_{2} = \frac{A-B}{1+A+B},
\end{equation}
where $A = \exp\big\{-\frac{\delta_{\mathrm{M+}}}{k_{\mathrm{B}}T_{\mathrm{H}}}\big\}$ and $B = \exp\big\{-\frac{\delta_{\mathrm{M-}}}{k_{\mathrm{B}}T_{\mathrm{C}}}\big\}$. Inserting the $T_{\mathrm{H}}=400$~K and $T_{\mathrm{C}}=100$~K into the equation we obtain an inversion of $I^{ideal}= 0.247$, which is slightly above the numerically calculated value of $I^{num}=0.244$. Hence, we can conclude that through the filters the heat gradient applied at the edge systems leads to an inversion in the middle system. 

The creation of an inversion depends sensitively on the energy filters as seen when introducing an energy mismatch $\Delta$  between the filter systems and the middle system in Fig.~\ref{fig:model_system}(c). Note that the mismatch is given such that the energy of state $|3\rangle^{(M)}$ increases to $\delta_{M+}+\Delta$. With increasing $\Delta$ the inversion decreases dramatically and for $\Delta > 8$~meV it changes its sign returning to a normal condition.

\begin{figure}[t]
	\centering
	\includegraphics[width=\columnwidth]{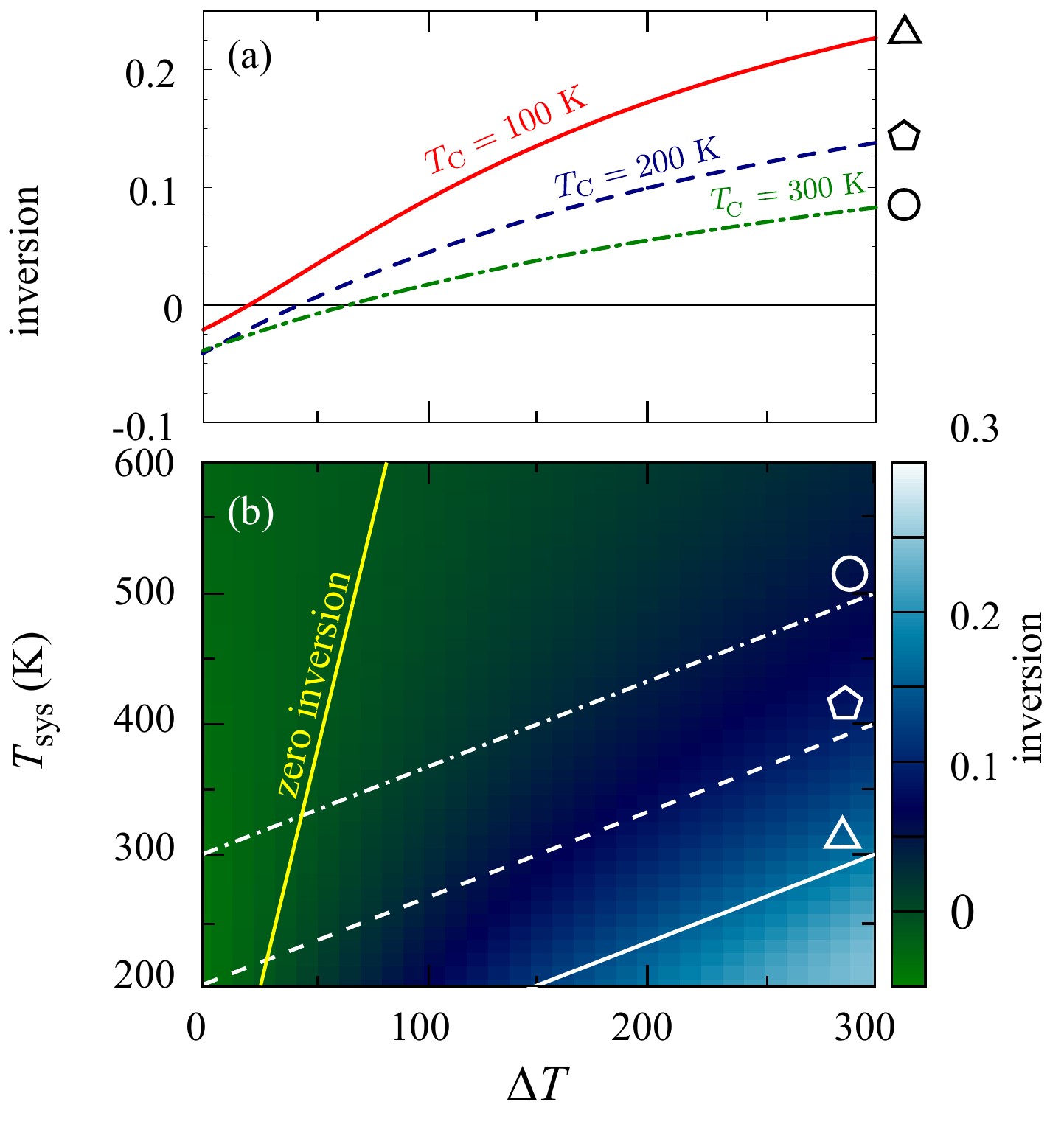}
	\caption{(a) Inversion as function of temperature difference $\Delta T$ for a constant temperature of the colder bath $T_{\mathrm{C}}=100$, $200$ and $300$~K. (b) Color map of the inversion as a function of temperature difference and the mean temperature of the system $T_{\mathrm{sys}} = (T_{\mathrm{H}}+T_{\mathrm{C}})/2$. The symbols (white circle, pentagon, and triangle) correspond to the curves from the upper panel. The yellow line depicts zero inversion.}
	\label{fig:inv_temps}
\end{figure}

Finding that a heat gradient is able to create a population inversion, given that appropriate filters are applied, we now consider the range of temperatures, for which a population inversion is possible in Fig.~\ref{fig:inv_temps}. Figure~\ref{fig:inv_temps}(a) shows the inversion, when keeping the temperature of the cold bath $T_{\mathrm{C}}$ fixed and increasing the temperature of the hot bath by $\Delta T$ to $T_{\mathrm{H}} = T_{\mathrm{C}}+ \Delta T$. As expected, for higher temperature differences the inversion increases, however, an inversion is only reached over a certain threshold given by the ratio $T_{\mathrm{H}}/T_{\mathrm{C}} > 1.2$. This threshold depends not only on the temperature difference, but also on the absolute values of the baths. This is also summarized  in Fig.~\ref{fig:inv_temps}(b), where we show a color plot of the inversion as a function of temperature difference of the heat baths (for the wider temperature range) and the mean temperature of the system defined as $T_{sys}=(T_{\mathrm{H}}+T_{\mathrm{C}})/2$. This figure underlines the fact, that a minimal temperature difference is needed to achieve inversion.

\begin{figure}[t]
	\centering
	\includegraphics[width=\columnwidth]{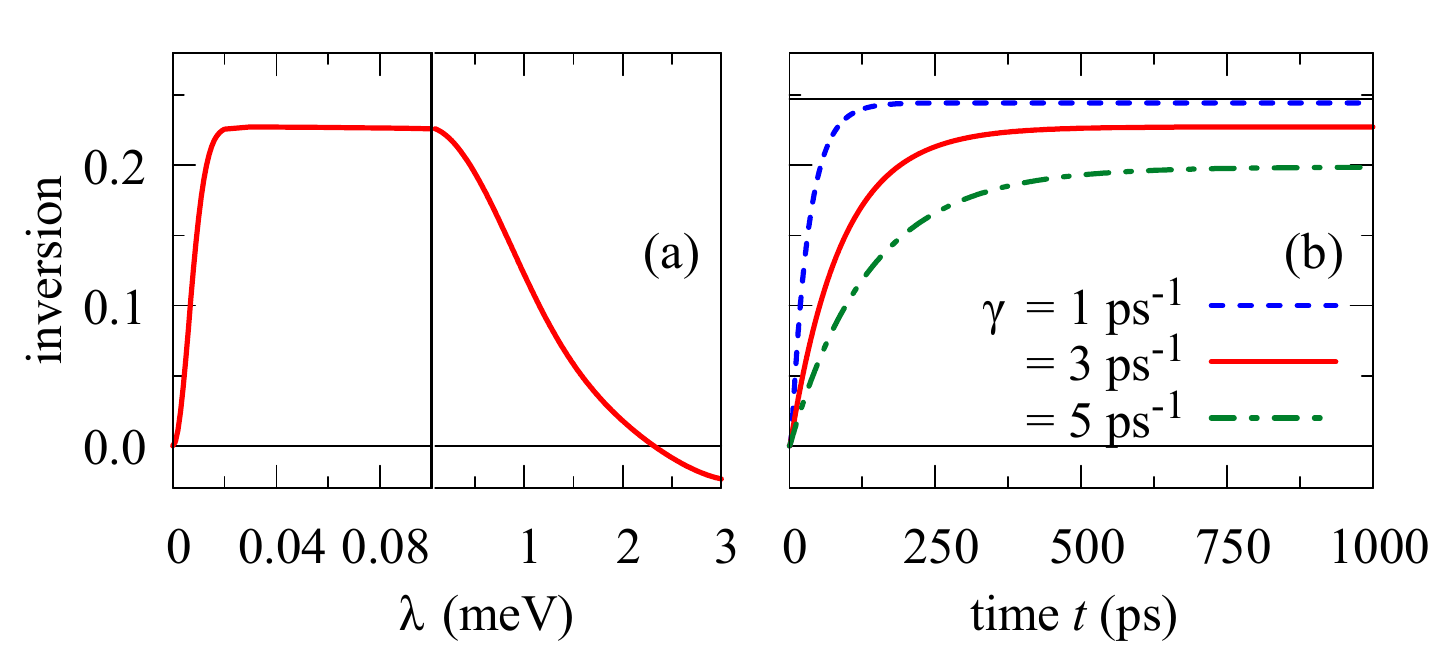}
	\caption{Stationary values of the inversion as function of (a) QS coupling parameter $\lambda$ and (b) heath bath coupling constant $\gamma$. The bath temperatures are $T_{\mathrm{H}}=400$~K and $T_{\mathrm{C}}=100$~K. The thin black line marks the ideal inversion.
	\label{fig:inv_lambda}
	}
\end{figure}

The inversion is also sensitive to other system parameters, like system coupling parameter $\lambda$ or the heat bath coupling parameter $\gamma$. This dependence is analyzed in Fig.~\ref{fig:inv_lambda}. In Fig.~\ref{fig:inv_lambda}(a) we show the stationary inversion as function of $\lambda$ keeping all other parameters fixed. The temperatures are $T_{\mathrm{H}}=400$~K and $T_{\mathrm{C}}=100$~K. For small $\lambda$ the inversion increases, because the efficient heat coupling is responsible to transfer the excitation to the center system. Then for a large range of $\lambda$ the inversion stays constant. For high values, the inversion decreases again, because now the interaction between the systems is not weak anymore and the new eigenstates become superposition of the uncoupled states. In Fig.~\ref{fig:inv_lambda}(b) we present the impact of the coupling of each two-level system to their environment on the inversion. The parameter $\gamma$ determines how fast the edge systems, and subsequently the middle system, are thermalised. With higher value of $\gamma$ the thermalisation becomes faster and as a consequence the inversion reaches nearly the ideal value of $I^{\mathrm{ideal}}$ marked by the black line (cf. Eq.~\eqref{eq:idealI}).

%
%The minimal temperature difference depends on the energy difference between the corresponding states and would increase drastically. This leads to the question, if our nanomaschine is also able to create photon lasing in the visible regime. The corresponding photon energy would lie in the order of a few eV. To achieve an inversion by a heat gradient, the temperature difference should be larger than XXXXX~K (10000?K), which is not possible to reach. In contrast, for THz photons, which have an energy range of a few meV, our nanomaschine would also work perfectly. 

%\textbf{Phonon - Theory} \\
Having seen that the heat gradient induces an inversion, we now study if this inversion can be utilized to drive a phonon laser. For the phonons, we assume a single acoustic phonon mode, which is confined by a cavity. Phonon cavities have been realized in semiconductor structures by superlattices \cite{trigo2002con,lanzillotti2007pho,kabuss2012opt,esmann2018top}. 

Denoting $b,b^{\dagger}$ as the bosonic operators of the phonon mode and $\omega$  its frequency, the phonon Hamiltonian is composed of the free phonon system and the carrier-phonon coupling
\begin{equation}
\hat{H}_{\mathrm{ph}} = \hbar \omega \hat{b}^{\dagger}\hat{b}+ \hat{H}_{\mathrm{c-ph}} \,.
\end{equation}
The energy of the phonon is chosen to be resonant with the energy difference in QS~M with $\hbar\omega=5$~meV. The phonon mode is coupled only to the middle QS M via
\begin{equation}
\hat{H}_{\mathrm{c-ph}} = \hat{1}^{(\mathrm{L})} \otimes \hat{h}^{(\mathrm{M})} \otimes \hat{1}^{(\mathrm{R})}
\end{equation}
with
\begin{equation}
\label{hm}
\hat{h}^{(\mathrm{M})} = \hbar g(\hat{b}^{\dagger}\hat{P}^{(\mathrm{M})}_{23} + \hat{b}\hat{P}^{(\mathrm{M})}_{32}).
\end{equation}
Here $g$ is the real coupling constant between the QS and phonon mode. We assume that the system can relax from the state $|3\rangle^{(\mathrm{M})}$ to state $|2\rangle^{(\mathrm{M})}$ by the emission of a phonon, while the reverse transition is possible by absorbing a phonon. 

A measurable quantity for the phonons is the lattice displacement, which is connected to the phonon operators via 
\begin{equation}
\langle  \hat{u} \rangle = u_{0} \left(\langle\hat{b}^{\dagger} \rangle +\langle \hat{b}^{} \rangle \right) =  u^{(+)} + u^{(-)},
\end{equation}
where we defined $u^{(+)} = u_{0}\langle\hat{b}\rangle$ and $u^{(-)} = u_{0}\langle\hat{b}^{\dagger}\rangle$ as well as the single phonon amplitude $u_{0}$. 

Introducing the phonon coupling, we extend the equations of motion to
\begin{equation}
\frac{d\hat{\rho}}{dt} = -\frac{i}{\hbar}[\hat{H}_{\mathrm{sys}}+\hat{H}_{\mathrm{int}}+\hat{H}_{\mathrm{ph}}  ,\hat{\rho}] + \hat{D}_{\mathrm{H}}(\hat{\rho}) + \hat{D}_{\mathrm{C}}(\hat{\rho}),
\end{equation}
For the lattice displacement this leads to the following rate equation
\begin{equation}
\frac{du}{dt} = -\Gamma u - iC \bigg(\rho_{23}^{(\mathrm{M})}(t) + \rho_{32}^{(\mathrm{M})}(t)\bigg),
\end{equation}
where we introduced the phonon dephasing rate $\Gamma$ and defined the coupling constant $C = u_{0}g$.
For our simulations we assume $\Gamma = 2$~ps$^{-1}$, $g = 2.25$~ps$^{-1}$, $u_{0} = 20$~pm. We further assume that always a very small, but finite displacement is present. 

\begin{figure}[t!]
	\centering
	\includegraphics[width=\columnwidth]{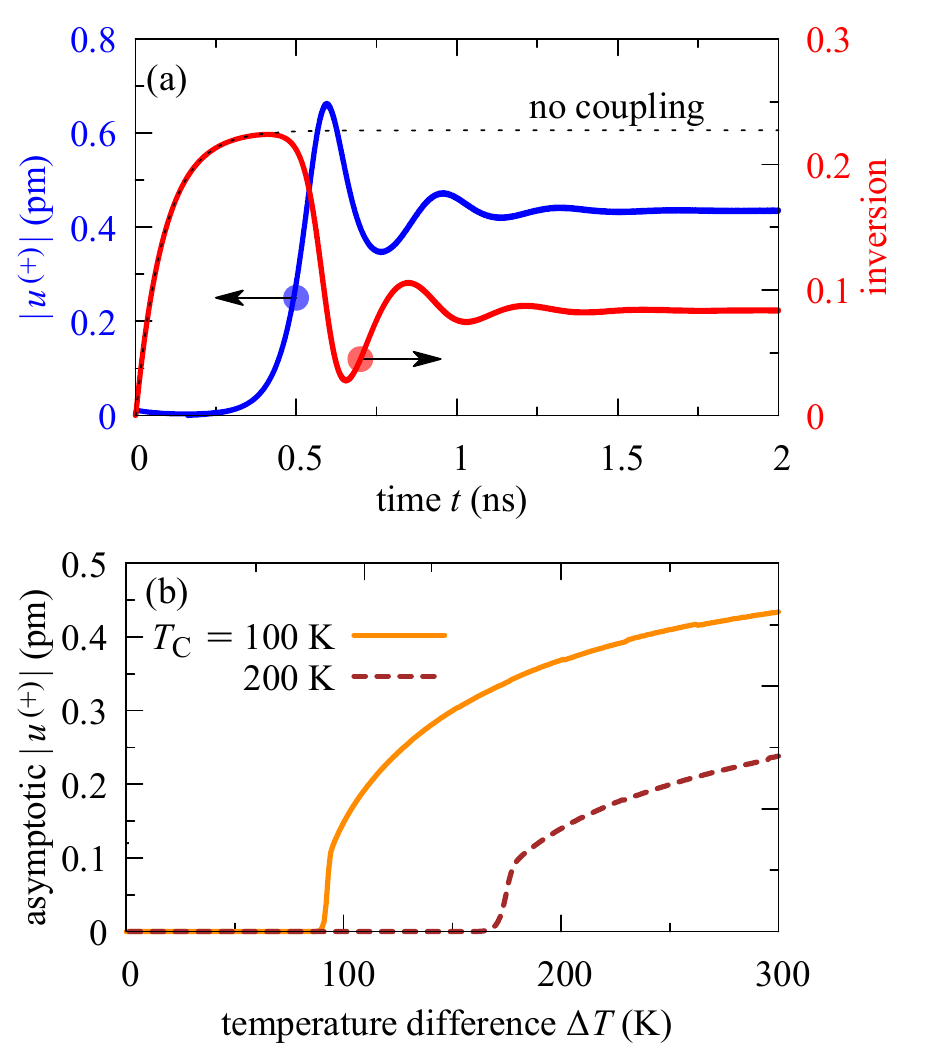}
	\caption{(a) Time evolution of the inversion (red) and the amplitude of the lattice displacement field (blue) for $T_{\mathrm{H}}=400$~K and $T_{\mathrm{C}}=100$~K.
	(b) Stationary amplitude of the lattice displacement field as function of temperature difference for $T_{\mathrm{C}}=100$~K (orange) and $T_{\mathrm{C}}=200$~K (brown).
		}
	\label{fig:lattice}
\end{figure}

%\textbf{Phonon lasing - Results}\\
In Fig.~\ref{fig:lattice}(a), we present the evolution of the inversion (red) in comparison to the amplitude of the lattice displacement field (blue). The temperatures are set to $T_{\mathrm{H}}=400$~K and $T_{\mathrm{C}}=100$~K. For small times the inversion of the system build up according to the thermalization of the middle QS~M. When the inversion is close to its maximum the lattice displacement starts to increase. Then the inversion and the lattice displacement oscillate against each other, which are relaxation oscillations well known from lasing dynamics. After some time a steady state is reached with a lattice displacement of $u_{\infty} =0.43$~pm. In this steady state a constant flow of coherent phonons is occurring, we therefore conclude that the system exhibits phonon lasing. 

In Fig.~\ref{fig:lattice}(b), we show the stationary amplitude as a function of the driving strength, which in our case is the temperature difference $\Delta T$. Keeping the cold bath fixed at $T_{\mathrm{C}}=100$~K (solid line), we increase the hot bath temperature. For a classical laser such an output curve should exhibit a characteristic onset of lasing at a given threshold  \cite{bjork1994definition}. Indeed, we find a characteristic onset of phonon-lasing at $\Delta T_{\mathrm{on}}=85$~K, where for temperature difference below $\Delta T_{\mathrm{on}}$ no phonons are emitted and the phonon amplitude rises significantly for temperature difference above $\Delta T_{\mathrm{on}}$ as expected for lasing. Like the inversion, the threshold depends not only on the temperature difference, but also on the absolute values of the temperature. If we fix the temperature of the cold bath to $T_{\mathrm{C}}=200$~K (dashed line), the threshold increases to $\Delta T'_{\mathrm{on}}=150$~K. 

%\textbf{Conclusions}\\
In conclusion, we have shown that a novel pumping mechanism using a thermal gradient is able to lead to inversion and to lasing. Our system was composed of a central three-level quantum system in interaction with quantum two-level subunits at each side acting as energy filters.  We then imposed a heat gradient on the whole system. Without any further interactions for sufficiently high temperature differences between the heat baths the system shows a positive inversion within the upper two levels of the central unit. That inversion could be utilized for the generation of coherent acoustic phonons in a phononic cavity. Our concept can readily be transferred to optomechanical systems resulting in phonon lasing in nanomechanical oscillators  \cite{kippenberg2007cavity,aspelmeyer2014cavity,anetsberger2009near}. Despite the opening of a strong heat conducting channel between the hot and the cold reservoirs due to the phonon-lasing action, the system shows an amplification of the lattice amplitude. Thus, the novel pumping mechanism using a thermal gradient is, indeed, able to produce coherent phonons demonstrating that amazingly a temperature difference can not only be used to drive a steam engine but also for the generation of coherent phonons in nanoscopic quantum systems.
%\section{APPENDIX}

%\bibliographystyle{prsty}

%%%\bibliography{abbr,quantum_DR}

%

\end{document}